# Domain tuning in mixed-phase BiFeO$_3$ thin films using vicinal substrates

Lu You,[1] Shintaro Yasui,[2] Xi Zou,[1] Hui Ding,[1] Zuhuang Chen,[1] Weigang Chen,[1] Lang Chen,[1] Hiroshi Funakubo,[2] Junling Wang[1,a)]

[1]*School of Materials Science and Engineering, Nanyang Technological University, Singapore 639798, Singapore*

[2]*Department of Innovative and Engineered Material, Tokyo Institute of Technology, Yokohama 226-8503, Japan*

[a)]Author to whom correspondence should be addressed. Electronic mail: jlwang@ntu.edu.sg.

# ABSTRACT

The structural and ferroelectric domain variants of highly-strained $BiFeO_3$ films grown on vicinal $LaSrAlO_4$ substrates were studied by piezoelectric force microscopy and high-resolution X-ray reciprocal space mapping. Through symmetry breaking of the substrate surface, ferroelastic domain variants in the highly-strained $M_C$ phase $BiFeO_3$ can be greatly reduced. Single-domain film can be obtained on substrates with large miscut angle, which is accompanied by the reduction of structural variants in the mixed-phase nanodomains. These findings lead to better understanding of the phase evolution and polarization rotation process in the strain-driven morphotropic phase system.

Multiferroic BiFeO$_3$ (BFO) film with a giant axial ratio (c/a ~ 1.23) can be stabilized under ultrahigh compressive strain (~4.5%) when grown epitaxially on LaAlO$_3$ substrate.[1,2] At small thickness (< 30 nm), the film adopts monoclinic M$_C$ structure at room temperature with the in-plane (IP) lattice distortion and polarization along <100> directions.[3,4] Such tetragonal-like M$_C$ phase is to be distinguished from the rhombohedral-like phases (including M$_A$ phase under small compressive strain, *M$_B$* phase under tensile strain and bulk rhombohedral phase), whose IP unit-cell distortion and polarization directions are along <110> axes.[5] With the increase of film thickness, rhombohedral-like phase with smaller c/a ratio emerges due to strain relaxation, which pairs with the tetragonal-like phase to form alternating stripe nanodomains embedded in the M$_C$ phase matrix. In such mixed phase nanodomains (MPNs), both the tetragonal-like and the rhombohedral-like phases are highly distorted with even lower symmetry compared to their parent counterparts.[6,7] Due to the proximate energy scale of these two phases, they are sensitive to external stimuli, leading to giant electromechanical response.[8]

However, the low-symmetry nature of the mixed phases also results in great complexity of the domain structures. Previous structural analysis revealed an eight-fold degeneracy of the mixed-phase variants, which is consistent with the observation from surface morphology.[6,9] In order to unravel the emergent properties of the morphotropic phase boundary (MPB),[10] it is highly desirable to reduce the domain variants in the $M_C$ phase matrix as well as the mixed-phase region. An effective method for domain engineering is through the control of substrate vicinality, which breaks the symmetry of the substrate surface so that specific ferroelastic domain variants are energetically favored. This approach has been widely exploited in the rhombohedral-like BFO films.[11,12] However, no such study has been reported on the highly-strained tetragonal-like BFO as well as the MPNs yet. In this report, we demonstrate that by growing BFO films on vicinal LaSrAlO$_4$ (LSAO) substrates, the ferroelastic variants of the $M_C$ phase matrix can be tailored. Furthermore, the structural variants of the MPNs can also be greatly reduced.

BFO films with thicknesses of 30 nm and 60 nm were deposited on (001)-oriented LSAO single crystal substrates with 0°, 2° and 4° miscut angles along [100] direction, respectively. Detailed film growth conditions can be found elsewhere.[13] The surface morphology and ferroelectric domain images were recorded using piezoelectric force microscopy (PFM) based on a commercial atomic force microscope (MFP-3D, Asylum Research). Structural analyses were performed on high-resolution X-ray diffractometer (PANalytical X'pert PRO MRD) using Cu K$_{\alpha 1}$ radiation ($\lambda$ = 1.5406 Å).

Figure 1 shows the surface morphologies together with the corresponding PFM images of the 30 nm films. Atomic-flat terraces can be observed in the no-miscut sample, while the 2°- and 4°-miscut samples exhibit step bunching feature which is typical for films grown on vicinal substrates.[14] These three samples all contain small amount of MPNs, which, however, do not affect the ferroelectric domain imaging of the $M_C$ matrix phase. The out-of-plane (OP) PFM images all show a uniform purple tone (-90° phase, insets of Fig. 1a-c), indicating complete downward-pointing polarization. All of the four ferroelastic variants that are allowed in $M_C$ symmetry can be found in the no-miscut sample. Every two pairs of the four variants can form stripe-like domain pattern with the domain boundaries running along <110> axes (Fig. 1d). Such domain structure is similar to that reported in the rhombohedral-like BFO films, except that the pattern is rotated by 45° in the film plane due to the symmetry change.[15] By increasing the substrate miscut angle to 2°, stripe-like domain pattern remains, however, with only three ferroelastic variants as shown in Figure 1e. Finally, on 4°-miscut substrate, we can only observe a predominant purple tone (the horizontal lines are artifacts caused by the large topography change at the step edges), which suggests a single-domain state with the IP polarization lying along the downhill miscut direction (Fig. 1f). This conclusion can be verified by scanning the sample again after 90° rotation. The resulting PFM image displays 0° phase signal, as the polarization vector is now parallel to the cantilever of the probe (data not shown). These results are qualitatively similar to rhombohedral-like BFO films grown on (001)-oriented $SrTiO_3$ substrate, where large miscut along [110] direction will lead to single domain variant, however, with the IP polarization opposite to the downhill miscut direction.[16] We suggest that this difference is due to the different strains experienced by the rhombohedral-like BFO on $SrTiO_3$ and the tetragonal-like BFO on LSAO. According to the first-principle calculations, the ground state of the "super-tetragonal" phase has an IP lattice constant around 3.66 Å,[17] smaller than that of LSAO (3.756 Å). And the OP lattice parameter is ~4.7 Å again smaller than the half unit-cell step of ~6.3 Å for LSAO. Therefore, tetragonal-like BFO grown on LSAO is in fact subjected to tensile strain,[13] which will induce a lattice distortion in favor of the observed polarization orientation (inset of Fig. 1f).[18]

With the reduced domain variants in $M_C$ matrix phase, we further studied the effect of substrate vicinality on the MPNs. The surface morphologies of 60 nm-thick films on 0°, 2° and 4° miscut substrates are displayed in Figure 2. Large amount of MPNs start to appear in all samples due to strain relaxation (less obvious in 4°-miscut sample because of the large height scale induced by step bunching). Different types of MPNs can be distinguished by defining structural vectors perpendicular to the stripe-like nanodomain boundaries.[19] As depicted in Figure 2a, totally eight types of MPNs coexist in no-miscut film, which suggests eightfold degeneracy in the film plane. The eight structural variants are summarized in Figure 2d, labeled from 1 to 4. The positive and negative superscripts are used to differentiate MPNs with opposite lattice tilting.[9] For example, variant $1^+$ and $1^-$ have the same orientation of the nanodomain boundaries, however, the tilting angle of the tetragonal-like and rhombohedral-like components are opposite as shown in the schematic of Figure 2d. With gradual increase of the miscut angle, the number of MPN variants appears to be reduced to 6 for 2°-miscut sample and only 2 for 4°-miscut sample. However, due to the significant step bunching in the 4°-miscut sample, the stripe-like features of some MPNs may be overwhelmed by the large height of the steps, especially those parallel to the bunching steps.

In order to further verify the structural variants existing in the films, plane-view X-ray reciprocal space mapping (RSM) was performed in the H-K plane. By fixing L value at the (001) *d*-spacing (4.21 Å) of the tilted rhombohedral-like phase in MPNs, we are able to resolve the domain variants in the film plane as shown in Figure 3. Clearly, eight diffraction spots can be observed for no-miscut sample, in good agreement with the result from topographic images. However, the 4°-miscut sample exhibits four peaks instead of two in the H-K plane (Fig. 3b). To explain this, we need to closely look into the polarization orientations in the MPNs. Based on our recent study,[9] the IP polarization of the tilted rhombohedral-like phase lies proximately to the <110> directions, while the tilted tetragonal-like phase in the MPNs has IP polarization directions rotating ~34° away from <100> directions. Thus, the polarization configurations in the MPNs can be mapped out as shown in Figure 2d (dash and solid arrows). Given that the $M_C$ phase matrix has a predominant downhill-pointing polarization, we can see the four structural variants in the right half of the coordinate ($1^+$, $2^+$, $3^-$, $4^-$) have the IP polarization component following the $M_C$ phase in the 4°-miscut sample. These are exactly the four variants observed in the RSM. According to recent temperature-dependent studies, there is a $M_C$-$M_A$ phase transition around 100 °C for the matrix phase.[20-24] Besides, the MPNs actually emerge from the matrix phase when cooling from high temperature.[25,26] By combining the above recent findings, we can envision the complicated phase evolution in the mixed-phase BFO films grown on 4°-miscut LSAO substrate during the cooling process as demonstrated in Figure 4. At high temperature, the film consists of highly-strained tetragonal-like $M_A$ phase with two possible polarization variants along <110> axes which follows the miscut direction. With deceasing temperature, MPNs start to appear, with the polarization vectors rotating in proximity around the $M_A$ matrix. Finally, the $M_A$ matrix phase transforms into $M_C$ phase, leaving the MPNs intact, as we have observed in the PFM and RSM images. Of course, further temperature-dependent study is required to reveal the detailed phase transition and polarization rotation process in the single-domain film.

In conclusion, domain structures in highly-strained BFO films with MPNs can be tailored through the substrate vicinality. The ferroelastic variants of the tetragonal-like $M_C$ can be reduced from four on no-miscut LSAO substrate to single domain on 4°-miscut substrate. Accompanying such domain selection is the reduction of types of the MPNs. The underpinning mechanism is closely linked to the thermodynamic phase evolution upon cooling. The mixed-phase BFO films with reduced structural variants will facilitate our further investigation on the multiferroic properties in this morphotropic phase system.

**Acknowledgements:** This work is supported by Nanyang Technological University and Ministry of Education of Singapore under projects ARC 16/08. Partial support from National Research Foundation of Singapore under project NRF-CRP5-2009-04 is also acknowledged.

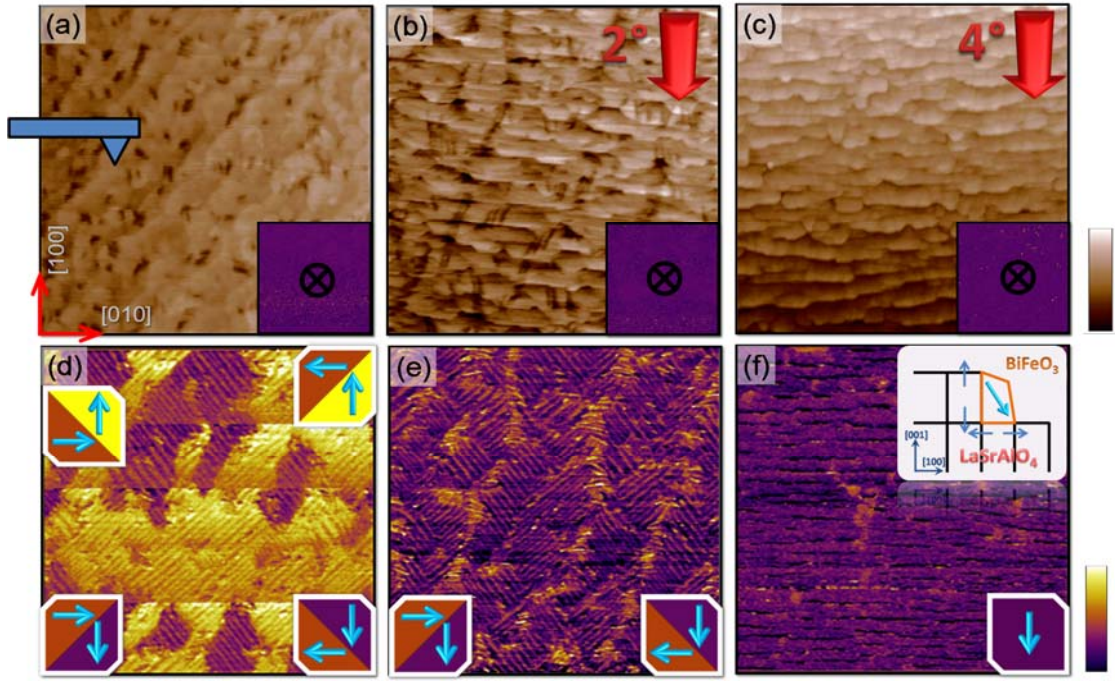

**Figure 1.** (Color online) (a), (b), (c) Surface morphology and (d), (e), (f) corresponding IP PFM image of 30 nm-thick BFO films grown on no-miscut, 2°-miscut and 4°-miscut LSAO substrates, respectively. The insets of (a), (b), (c) are the corresponding OP PFM images of three samples. The insets of (d), (e), (f) illustrate the polarization configurations. The schematic in (f) explains the preferred polarization direction induced by miscut susbstrate. All images are 3 μm * 3 μm. The height scale in (a), (b), (c) is 5 nm, 5 nm, 30 nm, respectively. The phase scale for PFM images are all 180°.

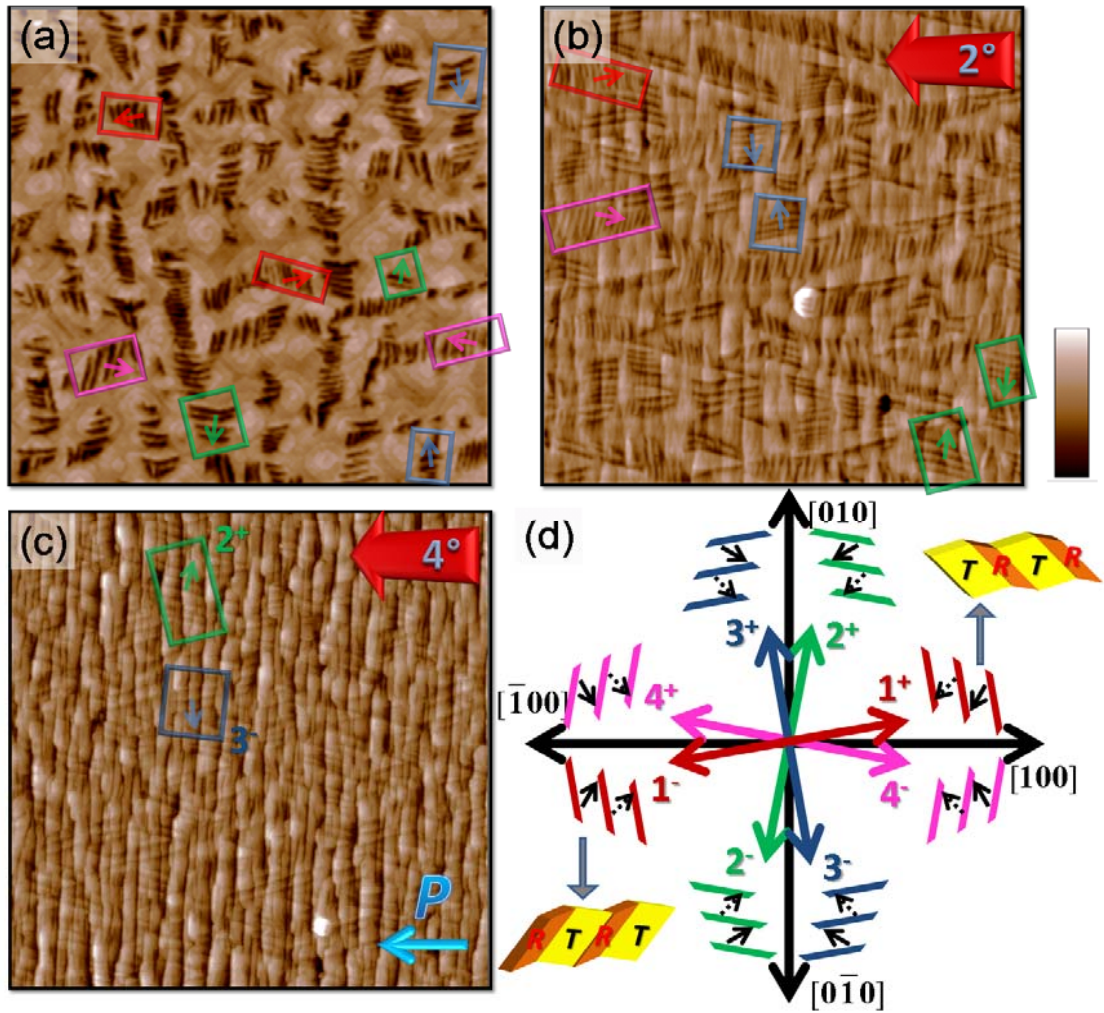

**Figure 2.** (Color online) Surface morphology of 60 nm-thick BFO thin films grown on (a) no-miscut, (b) 2°-miscut and (c) 4°-miscut LSAO substrates. (d) The eight possible variants of MPNs denoted by the structural vectors. The polarization configuration in each mixed-phase array is indicated by black arrow. Perspective view of the $1^+$ and $1^-$ variants are also shown. The different types of MPNs in (a), (b) and (c) are identified by arrows according to (d). All images are 3 μm * 3 μm. The height scale in (a), (b), (c) is 5 nm, 8 nm, 15 nm, respectively.

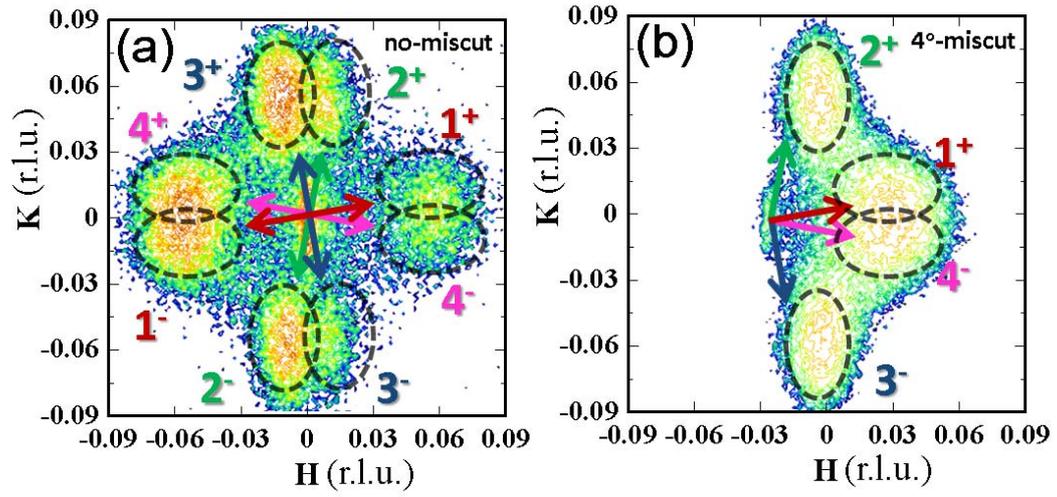

**Figure 3.** (Color online) (001) H-K RSM of the rhombohedral-like phase in the MPNs of (a) no-miscut and (b) 4°-miscut sample. The diffraction peaks are assigned to corresponding structural vectors by arrows. r.l.u. = reciprocal lattice unit.

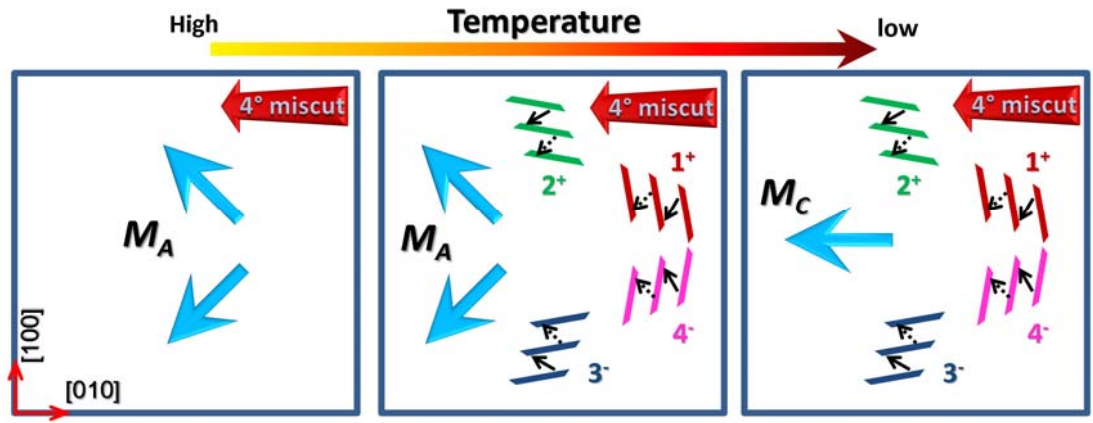

**Figure 4.** (Color online) Temperature-dependent phase evolution in the highly-strain BFO thin film with MPNs grown on 4°-miscut LSAO substrate.